\documentclass[aps,prl,twocolumn]{revtex4}
\usepackage{amsmath}
\include{psfig}
\input epsf

\newcommand{\PIS}{P_\text{IS}}
\newcommand{\EIS}{E_\text{IS}}
\newcommand{\Fvib}{F_\text{vib}}
\newcommand{\Sconf}{S_\text{conf}}
\newcommand{\Te}{T_\text{e}}

\begin{document}

\title{A test of non-equilibrium thermodynamics in glassy systems: the soft-sphere case}
\author{E. La Nave$^{*}$, F. Sciortino$^{*}$, P. Tartaglia$^{*}$}
 \address{ $^{*}$Dipartimento di Fisica, INFM UdR and Center for Statistical Mechanics and Complexity, 
Universit\`{a} di Roma La Sapienza, P.le A. Moro 5, I-00185 Rome, Italy}
\author{M. S. Shell$^{\dag}$ and P. G. Debenedetti$^{\dag}$}
\address{$^{\dag}$Department of Chemical Engineering, Princeton University, Princeton, NJ 08544}

\begin{abstract}
The scaling properties of the soft-sphere potential
allow the derivation of an exact expression for the pressure 
of a frozen liquid, i.e., the pressure corresponding to configurations which are local minima in its multidimensional potential energy landscape.  
The existence of such a relation offers 
the unique possibility for testing the recently proposed
extension of the liquid free energy
to glassy out-of-equilibrium conditions and the 
associated expression for the temperature of the 
configurational degrees of freedom.  We demonstrate
that the non-equilibrium free energy provides an exact
description of the soft-sphere pressure in glass states.
 \end{abstract}

\maketitle

The potential energy landscape (PEL) formalism \cite{sw} 
has provided a transparent formulation of
the equilibrium free energy of supercooled liquids 
based on the statistical properties of a system's 
multidimensional potential energy surface \cite{sw,skt99,sastry01,speedyjpc,heuer,lewis,eos,scott}, i.e.,
in terms of the number, depth, and associated phase-space volume of 
local potential energy minima \cite{basin}.  
This formalism, in which the PEL minima are termed
inherent structures (IS), is well suited to the description
of supercooled liquids and glasses.  While the liquid 
explores an exponentially large number of distinct PEL basins, a glass
explores only a very small fraction of these (within an experimentally-accessible time frame.)
The equilibrium free energy is written as sum of two contributions: an
entropic term, $-T\Sconf(\EIS)$, which accounts
for the number of basins of depth $\EIS$, and    
the term $\EIS+\Fvib(T,\EIS)$, which describes the free energy of 
the system confined to an average basin of depth $\EIS$.
Here, $\Sconf$ is associated with the energy degeneracy of
mechanically stable configurations and hence is termed 
the configurational entropy, while $\Fvib$ is associated
with the ``vibrational,'' kinetic distortions of the system around
these configurations.  Analogous expressions have been derived 
from alternative theoretical approaches
\cite{parisimezard,tap,teo,shultz98}.

The extension of the supercooled liquid free-energy
to the out-of-equilibrium case has recently been
proposed \cite{kurchan,teo,leuzzi,franz,st01,parrocchia}. 
In the PEL formalism, out-of-equilibrium conditions are implemented 
by imposing the constraint that the basin explored by the system
while aging does not coincide with the typical equilibrium one \cite{st01}.  
The proposed free energy is given by

\begin{equation}
F(V,T,\Te) = - \Te \Sconf(V,\EIS) + \EIS +\Fvib(V,T, \EIS)
\label{eq:freee}
\end{equation}

The main distinction between this expression and the equilibrium case is
that while $\Fvib$ is still evaluated at the thermostat temperature $T$, the
configurational entropy term is weighted by an additional temperature $\Te$, 
which may be thought of as the temperature of the 
(out-of-equilibrium) configurational degrees of freedom. 
When $\Te$ and $T$ differ, the system is in a non-equilibrium state.
While in equilibrium the value of $\EIS$ is
controlled only by $T$ and $V$ via the condition 
$\partial F(T,V)/\partial \EIS=0$, in this non-equilibrium setting the
value of $\EIS$ is also a function of $\Te$.  In this case
the basin depth which the system samples, $\EIS(V,T,\Te)$,
is the solution of 

\begin{equation}
1+ {{ \partial \Fvib }\over { \partial \EIS }}-
\Te
{ {\partial \Sconf } \over {\partial \EIS} }=0
\label{eq:tint}
\end{equation}
\noindent

This expression is based on the hypothesis that the out-of-equilibrium 
system samples a distribution of basins similar to the one explored in
equilibrium \cite{franz,kurchan,st01,leuzzi}.
Inverting $\EIS(V,T,\Te)$ provides an estimate of
$\Te$ when the bath temperature is $T$ and the
system is confined to a basin of depth $\EIS$.
As already discussed \cite{st01}, the expression for 
$\Te$ coincides with the experimentally determined 
fictive temperature \cite{tool} for models in which
$\partial \Fvib/\partial \EIS=0$, i.e., when the 
phase-space volume of basins is independent of
their depth. The same expression for $\Te$ 
has been derived also by Franz and 
Virasoro \cite{franz} in the context of $p-$spin systems, 
once the basin free energy is identified with the TAP free
energy \cite{tap}.

Starting from the proposed free energy (Eq. \ref{eq:freee})
it is possible to calculate the thermodynamic pressure
in out-of-equilibrium conditions, when the
bath temperature is $T$ and the configurational temperature
is $\Te$ (i.e. when the system is exploring basins different from
those explored in equilibrium at temperature $T$). 
To this end, we evaluate the constant-$T$ volume-derivative of the 
free energy in Eq. \ref{eq:freee},
\begin{widetext}
\begin{equation}
P(V,T,\Te)= -\left(\frac{\partial F(V,T,\Te)}{\partial V}\right)_{T,\Te}
\end{equation}
\begin{equation}
=\Te \left[ \left( \frac {\partial \Sconf}{\partial \EIS} \right)_{V}\left( \frac {\partial \EIS}{\partial V} \right)_{T,\Te} + \left( \frac {\partial \Sconf}{\partial V} \right)_{\EIS} \right]-\left(\frac {\partial \EIS}{\partial V}\right)_{T,\Te}-
 \left( \frac {\partial \Fvib}{\partial \EIS} \right)_{T,V}\left( \frac {\partial \EIS}{\partial V} \right)_{T,\Te} -
\left(\frac {\partial \Fvib}{\partial V}\right)_{T,\EIS},
\end{equation}
and rearranging:
\begin{equation}
P(V,T,\Te)=\left( \frac {\partial \EIS}{\partial V} \right)_{T,\Te}
 \left[ \Te \left( \frac {\partial \Sconf}{\partial \EIS} \right)_{V}-1-  \left( \frac {\partial \Fvib}{\partial \EIS} \right)_{T,V}  \right] + \Te \left( \frac {\partial \Sconf}{\partial V} \right)_{\EIS} - \left(\frac 
{\partial \Fvib}{\partial V}\right)_{T,\EIS}.
\label{eq:pout}
\end{equation}
\end{widetext}
The first term in the rhs of Eq.~\ref{eq:pout} is zero by
the $\EIS$ condition in Eq.~\ref{eq:tint}.  Thus,
\begin{equation}
P(V,T,\Te)= \Te \left( \frac {\partial \Sconf }{\partial V} \right)_{\EIS} - \left(\frac{\partial \Fvib}{\partial V}\right)_{T,\EIS}.
\end{equation}

The above expression, when evaluated at the bath temperature 
$T=0 K$, provides the theoretical expression
for the pressure experienced in an inherent structure, $\PIS$. 
Indeed, an infinite cooling rate quench to $T=0 K$ brings the system 
to the local inherent structure \cite{sw,fn:quench}.  In other words, the steepest descent 
procedure used numerically to locate an inherent structure is equivalent
to the physical process of setting the bath temperature to $T=0 K$.
The absolute zero bath temperature eliminates any contribution
arising from the vibrational free energy and hence, $\PIS(V,\EIS)$ is 
\begin{equation}
\PIS(V,\EIS)=P(V,0,\Te)=  \Te \left( \frac {\partial \Sconf }{\partial V} \right)_{\EIS}. 
\label{eq:pistermo}
\end{equation}
Using the mathematical identity
\begin{equation}
 \left( \frac {\partial \Sconf }{\partial V}       \right)_{\EIS}=-
 \left( \frac {\partial \Sconf }{\partial \EIS} \right)_{V}
 \left( \frac {\partial \EIS        }{\partial V }        \right)_{\Sconf},
\label{cyclic}
\end{equation}
and the $\EIS$ condition in Eq.~\ref{eq:tint} evaluated at $T=0$, we obtain
\begin{eqnarray}
\PIS(V,\EIS) & = &
- \Te 
\left( \frac {\partial \Sconf }{\partial \EIS} \right)_{V}
 \left( \frac {\partial \EIS        }{\partial V }        \right)_{\Sconf}
\nonumber \\
& = & -\left( \frac {\partial \EIS}{\partial V }       \right)_{\Sconf}
\label{eq:pistermo2}
\end{eqnarray}
which provides an alternative definition of the inherent structure pressure.

We now show that the scaling properties of the soft-sphere
potential provides a consistency check for the derived expression
for $\PIS$ and, simultaneously, the validity of Eqs.~\ref{eq:freee} and \ref{eq:tint}.
Indeed, the thermodynamic expression for $\PIS$ is explicit in the case of the
soft-sphere potential, without further assumptions.
This potential has been extensively studied as a
model for liquids and glasses \cite{fn,hansen,wolynes,parisiprl,coluzzimezard,coluzzi,deben99,still01,scott,speedypisa}. 
The potential energy $E$ of a system composed of $N$ particles interacting 
via a soft-sphere potential is $E({\bf r}^N)=\sum_{i,j>i=1}^{N} 
\epsilon \left(\sigma/|{\bf r}_i - {\bf r}_j| \right)^n$,  
where ${\bf r}^N=\{{\bf r}_1,{\bf r}_2,...., {\bf r}_N\}$ with
${\bf r}_i$ the coordinates of particle $i$, and $\epsilon$ and
$\sigma$ fix the energy and length scales, respectively. 

The self-similar nature of the soft-sphere potential has a remarkable 
property in that uniform scaling of particle coordinates does not 
produce changes the topology of the potential energy landscape, 
since $E(\lambda {\bf r^N})=\lambda^{-n} E({\bf r^N})$ \cite{still99-s,still01-s}.  
In the PEL formalism this scaling property implies that the total 
number of inherent structures and basins is invariant to volume 
changes.  The scaling has two important additional consequences 
which we exploit in the present study: (i) an isotropic compression 
of a configuration which is a local potential energy minimum remains 
a local minimum; the potential energy change associated with the 
compression coincides with the change in the $\EIS$ value; 
(ii) an isotropic compression moves the soft-sphere system along a path of
constant configurational entropy (since the number of basins of 
depth $\EIS$ at volume $V$ is identical to the number of basins 
of depth $\EIS+\delta \EIS$ at volume $V+\delta V$).
Because $P_{IS}$ is  the measure of the change of  system's energy under
 a compression of an $IS-$configuration, these two considerations 
allow us to write  $\PIS$  
as the volume derivative of the inherent structure energy along a constant
configurational entropy path:
\begin{equation}
\PIS(V,\EIS)=  - \left( \frac {\partial \EIS}{\partial V }  \right)_{\Sconf}.
\label{eq:pissoft}
\end{equation}
\noindent
This expression, whose derivation has been based only on the self-similar nature
of the soft-sphere potential, coincides with the general expression in 
Eq.~\ref{eq:pistermo2} derived from the out-of-equilibrium thermodynamic approach
and provides a strong validation of the proposed Eqs.~\ref{eq:freee}-\ref{eq:tint}.

To the extent that the out-of-equilibrium formalism  is an appropriate
description of systems beyond soft-spheres, the expression for
the inherent structure pressure given by Eq.~\ref{eq:pistermo2} 
is quite general.  Encouraging support for this statement is given by
the behavior of the inherent structure pressure below an ideal glass
transition\cite{comment} . 

To summarize, the present Letter provides support for the 
recently proposed out-of-equilibrium approach to supercooled
liquids and its corresponding definition of configurational
temperature (i.e., the temperature characterizing a system's
sampling of inherent structures which are distinct from those 
it would sample in equilibrium.)  In the context of the
soft-sphere system, this non-equilibrium formalism provides
a consistent expression for the inherent structure pressure,
$\PIS$.  Furthermore, the results presented in this Letter provide a 
formal and general derivation of $\PIS$ (Eq.~\ref{eq:pistermo2})
in terms of statistical properties of the landscape 
\cite{parrocchia,scott} and open the way for a consistent 
formulation of thermodynamic properties in disordered materials
based on a separation of configurational and vibrational properties.

The Roma group acknowledges support from Miur COFIN 2002 and FIRB and INFM-PRA GenFdT.
MSS gratefully acknowledges the support of the Fannie and John Hertz Foundation.  
PGD gratefully acknowledges the support of the Department of Energy, Division of Chemical Sciences, Geosciences, and Biosciences, Office of Basic Energy Science (grant DE-FG02-87ER13714.)  We thank J. Kurchan, L. Leuzzi  and G. Parisi for constructive criticisms.



\begin{references}
\bibitem{basin}
A basin is properly defined as the set of points in configuration space which map to the same local minima under a steepest 
descent path.
\bibitem{fn:quench}
We consider a quench of infinite cooling rate to be 
the continuous depletion of kinetic energy; 
as a result of this procedure, the system relaxes to its
local mechanically stable configuration.
\bibitem{fn} Note that 
the excess free energy of the soft-sphere 
system is a function of only one thermodynamic variable
$z=\sigma^3
 \frac {(\beta \epsilon)^{3/n}} {V}$ \cite{hansen}.
\bibitem{comment} Intuitively, configurational properties should 
be temperature-independent at temperatures below an ideal glass transition, 
where the system is confined to a configurationally unique structure.  
We see from Eq.~\ref{eq:pistermo2} that if the average inherent structure
energy is $T$-independent and the system is
in a unique configuration ($\Sconf=0$), the inherent structure
pressure is indeed also $T$-independent\protect\cite{scott}.
\end{references}
\end{document}